# Fermilab Antiproton Source, Recycler Ring, and Main Injector

Sergei Nagaitsev, Fermilab, Batavia, IL 60510

**Introduction**

The antiproton source for a proton-antiproton collider at Fermilab was proposed in 1976 [1]. The proposal argued that the requisite luminosity (~$10^{29}$ cm$^{-2}$sec$^{-1}$) could be achieved with a facility that would produce and cool approximately $10^{11}$ antiprotons per day. Funding for the Tevatron I project (to construct the Antiproton source) was initiated in 1981 and the Tevatron ring itself was completed, as a fixed target accelerator, in the summer of 1983 and the Antiproton Source was completed in 1985. At the end of its operations in 2011, the Fermilab antiproton production complex consisted of a sophisticated target system, three 8-GeV storage rings (namely the Debuncher, Accumulator and Recycler), 25 independent multi-GHz stochastic cooling systems, the world's only relativistic electron cooling system and a team of technical experts equal to none. Sustained accumulation of antiprotons was possible at the rate of greater than $2.5 \times 10^{11}$ per hour. Record-size stacks of antiprotons in excess of $3 \times 10^{12}$ were accumulated in the Accumulator ring and $6 \times 10^{12}$ in the Recycler. In some special cases, the antiprotons were stored in rings for more than 50 days. Note, that over the years, some $10^{16}$ antiprotons were produced and accumulated at Fermilab, which is about 17 nanograms and more than 90% of the world's total man-made quantity of nuclear antimatter.

The accelerator complex at Fermilab supported a broad physics program including the Tevatron Collider Run II [2], neutrino experiments using 8 GeV and 120 GeV proton beams, as well as a test beam facility and other fixed target experiments using 120 GeV primary proton beams. The following sections provide a brief description of Fermilab accelerators as they operated at the end of the Collider Run II (2011).

### 1. The Proton Source

The Proton Source consists of the Pre-Accelerator (Pre-Acc), the Linac, and the Booster. For operational redundancy, there were two independent 750-kV Pre-Acc systems which provided H$^-$ ions for acceleration through the Linac. Each Pre-Acc was a Cockcroft-Walton accelerator having its own magnetron-type H$^-$ source running at a 15 Hz repetition rate, a voltage multiplier to generate the 750 kV accelerating voltage, and a chopper to set the beam pulse length going into the Linac. The typical H$^-$ source output current is 40-60 mA.

The Linac accelerated H$^-$ ions from 750 keV to 400 MeV. Originally, the Linac was a 200 MeV machine made entirely of Alvarez-style drift tube tanks [3], but a 1991 upgrade replaced four of the drift tubes with side coupled cavities to allow acceleration up to 400 MeV [4]. Today, the low energy section (up to 116 MeV) is made of drift tube tanks operating with 201.25 MHz RF fed from triod-based 5 MW power amplifier tubes. The high energy section (116 – 400 MeV) consists of 7 side-coupled cavity girders powered by 805-MHz, 12-MW Klystrons providing a gradient of ≈ 7 MV/m. A transition section between the two linac sections provides the optics matching and rebunching into the higher frequency RF system. The nominal beam current in the Linac is 30-35 mA.

The Booster is a 474 meter circumference, rapid-cycling synchrotron ramping from 400 MeV to 8 GeV at a 15 Hz repetition rate. (Note that while the magnets ramp at 15 Hz, beam is not present on every cycle.) Multi-turn injection is achieved by passing the incoming H$^-$ ions through 1.5 µm thick (300 µg/cm$^2$) carbon stripping foils as they merge with the circulating proton beam on a common orbit. The 96 10-foot long combined-function Booster gradient magnets are grouped into 24 identical periods in a FOFDOOD lattice [5]. The Booster RF system (harmonic number = 84) consists of 19 cavities (18 operational + 1 spare) that must sweep from 37.9-52.8 MHz as the beam velocity increases during acceleration. The ferrite tuners and power amplifiers are mounted on the cavities in the tunnel. The cavities provide a total of ≈750 kV per turn for acceleration. The Booster transition energy (4.2 GeV) occurs at 17 ms in the cycle. The Booster throughput efficiency is 85-90% for typical beam intensities of 4.5-5.0×10$^{12}$ protons per pulse. The majority of the proton flux through the Booster is delivered to the 8 GeV and 120 GeV neutrino production targets.

## 2. The Main Injector

The Main Injector [6] (MI) is a 3319.4-m circumference synchrotron, which can accelerate both proton and antiproton beams from 8 GeV up to 150 GeV. It has a FODO lattice using conventional, separated function dipole and quadrupole magnets. There are also trim dipole and quadrupoles, skew quadrupole, sextupole and octupole magnets in the lattice. Since the Main Injector circumference is seven times the Booster circumference, beam from multiple consecutive Booster cycles, called batches, can be injected around the Main Injector. In addition, even higher beam intensity can be accelerated by injecting more than seven Booster batches through the process of slip-stacking: capturing one set of injected proton batches with one RF system, decelerating them slightly, then capturing another set of proton injections with another independent RF system, and merging them prior to acceleration. There are 18 53-MHz RF cavities, grouped into 2 independently controlled systems to allow slip-stacking and the flexibility for maintenance. Beam-loading compensation and active damping systems have been implemented to help maintain beam stability. For beam injections into the Tevatron, coalescing of several 53 MHz bunches of protons and antiprotons into single, high intensity bunches also requires a 2.5 MHz system for bunch rotations and a 106 MHz cavity to flatten the potential when recapturing beam into the single 53 MHz bunch to be injected into the Tevatron ring. A set of collimators was installed in the Main Injector to help localize beam losses to reduce widespread activation of ring components in the tunnel.

The Main Injector supports various operational modes for delivering beam across the complex. For the antiproton and neutrino production, up to 11 proton batches from the Booster were injected and slip-stacked prior to acceleration. After reaching 120 GeV, 2 batches were extracted to the antiproton production target while the remaining 9 batches were extracted to the NuMI neutrino production target (Figure 1). At its peak performance, the Main Injector can sustain 400 kW delivery of 120 GeV proton beam power at 2.2 sec cycle times. The Main Injector also provides 120 GeV protons in a 4 sec long slow-spill extracted to the Switchyard as a primary beam or for production of secondary and tertiary beams for the Meson Test Beam Facility and other fixed-target experiments. In addition, the Main Injector served as an "effective" transport line for 8 GeV antiprotons being transferred from the Accumulator to the Recycler for later use in the Tevatron. Protons from the Booster and antiprotons from the Recycler were accelerated to 150 GeV in the Main Injector and coalesced into higher intensity bunches for injection into the Tevatron for a colliding beam store.

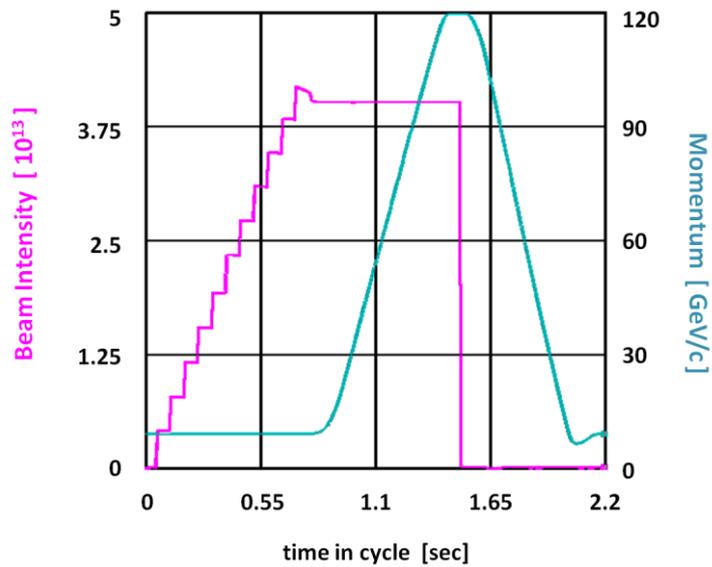

**Figure 1:** The Main Injector cycle illustrating an 11-batch proton injection, acceleration and extraction.

### 3. The Antiproton Source

The Antiproton Source [7] had 3 main parts: the Target Station, the Debuncher, and the Accumulator. Each of these is described briefly below while outlining the steps of an antiproton production cycle. In the Target Station, batches of 120-GeV protons ($\sim 8\times 10^{12}$ per batch), delivered from the Main Injector, strike the Inconel (a nickel-iron alloy) target every 2.2 sec. The beam spot on the target can be controlled by a set of quadrupole magnets. The target is rotated between beam pulses to spread the depletion and damage uniformly around its circumference. The shower of secondary particles, emanating from the target, was focused both horizontally and vertically by a pulsed, high current lithium lens that can provide up 1000 T/m gradient.

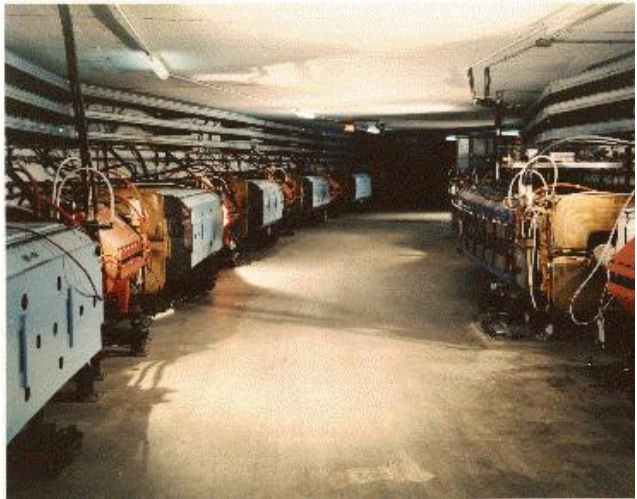

**Figure 2:** The Antiproton Source consists of the Debuncher ring (outer circumference, left) and the Accumulator ring (inner circumference, right).

Downstream of the Li lens was a pulsed dipole magnet, which steered negatively-charged particles at 8.9-GeV/c momentum into the transport line toward the Debuncher. A collimator between the lens and pulsed magnet was installed to help protect the pulsed magnet from radiation damage as the incoming primary proton beam intensity increased with proton slip-stacking in Main Injector.

The Debuncher and Accumulator (Figure 2) were both triangular-shaped rings of conventional magnets sharing the same tunnel. While the Debuncher had a FODO lattice, the Accumulator lattice had particular features needed for cooling and accumulating antiprotons with stochastic cooling systems. A total of 21 independent stochastic cooling systems were implemented in the Accumulator and Debuncher [8]. Such a variety of cooling systems was possible after a series of development efforts [9, 10] allowing for more robust and less expensive pick-up arrays.

The $\sim 2 \times 10^8$ bunches of antiprotons entering the Debuncher from the transport line retained the 53-MHz bunch structure from the primary proton beam on the target. A 53-MHz RF system (harmonic number = 90) was used for the bunch rotation and debunching of the antiprotons into a continuous beam with a low momentum spread. An independent 2.4 MHz RF system provided a barrier bucket to allow a gap for extraction to the Accumulator. Stochastic cooling systems reduced the transverse emittance from 300 to 30 π mm-mrad (rms, normalized) and momentum spread from 0.30% to <0.14% prior to injection into the Accumulator.

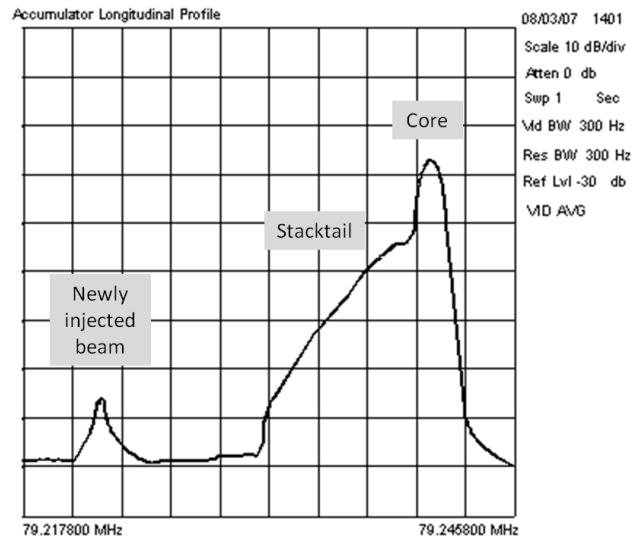

**Figure 3:** The frequency (energy) distribution of antiprotons in the Accumulator highlighting incoming antiprotons (left), the stacktail beam (middle) being cooled and decelerated toward the core (right). Higher beam energy is to the left, lower energy is to the right.

In the Accumulator, antiprotons were momentum-stacked and cooled by a series of RF manipulations and stochastic cooling. The incoming antiprotons were captured and decelerated by 60 MeV with a 53-MHz RF system (harmonic number = 84) to the central orbit where the beam was adiabatically debunched. Before the next pulse of antiprotons arrived (every 2.2 sec), the so-called stacktail momentum stochastic cooling system [11] decelerated the antiprotons another 150 MeV toward the core orbit where another set of independent betatron and momentum stochastic cooling systems provided additional cooling while building a "stack" of antiprotons. Figure 3 illustrates the frequency (energy) distribution of antiprotons in the Accumulator. Figure 4 shows the average antiproton accumulation rates since 1994; typical values at the end of Run II were in the range 24-26 $\times 10^{10}$/hr, with a maximum recorder rate of 28.5 $\times 10^{10}$/hr.

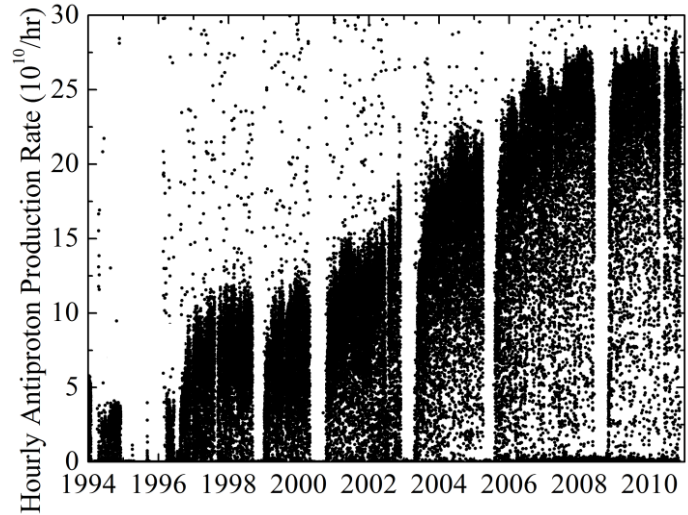

**Figure 4:** The average antiproton accumulation rate since 1994 and during all of Collider Run II (including the production in the Antiproton Source and storage in the Recycler). ). Note that some data points at the highest rates, in particular in the early years, are merely artifacts of the data acquisition and logging system.

### 3. The Recycler

The Recycler [12] is a permanent-magnet, fixed momentum (8.9-GeV/c) storage ring located in the Main Injector tunnel (Figure 5). As conceived the Recycler would provide storage for very large numbers of antiprotons (up to $6\times10^{12}$) and would increase the effective production rate by recapturing unused antiprotons at the end of collider stores (hence the name Recycler). Recycling of antiprotons was determined to be ineffective and was never implemented. However, the Recycler was used as a final antiproton cooling and storage ring for accumulating significantly larger stashes (so called to differentiate from Accumulator 'stacks') of antiprotons than can be accumulated in the Antiproton Accumulator. The main Recycler magnets are combined-function strontium ferrite permanent magnets arranged in a FODO lattice. Trim electro-magnets are used to provide orbit and lattice corrections.

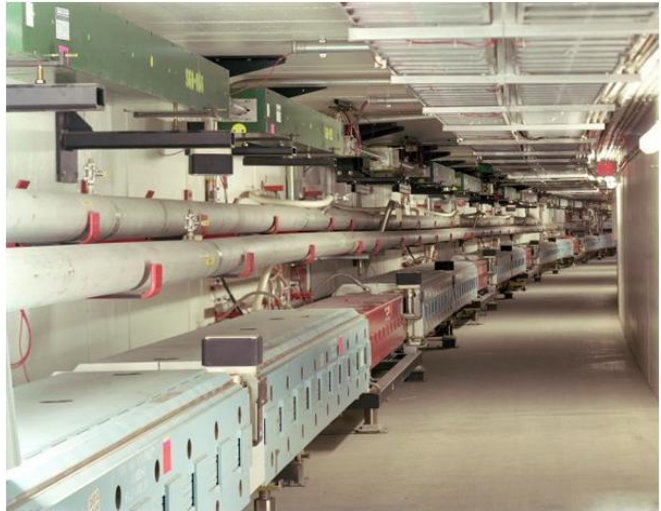

**Figure 5:** The Recycler (top) and the Main Injector (bottom) rings installed in a common tunnel.

An important feature of the Recycler was an electron cooling system [13]. It augmented the Recycler's cooling capability and complemented the stochastic cooling system and its inherent limitations. The Pelletron (an electrostatic accelerator manufactured by the National Electrostatics Corp.) provided a 4.3 MeV electron beam (up to 500 mA) which overlaped with the 8-GeV antiprotons, circulating in the Recycler, in a 20-m long section and cooled the antiprotons both transversely and longitudinally. Figure 6 shows the schematic layout of the Fermilab electron cooling system. The dc electron beam was generated by a thermionic gun, located in the high-voltage terminal of the electrostatic accelerator. This accelerator was incapable of sustaining dc beam currents to ground in excess of about 100 μA. Hence, to attain the electron dc current of 500 mA, a recirculation scheme was employed. A typical relative beam current loss in such a scheme was $2 \times 10^{-5}$.

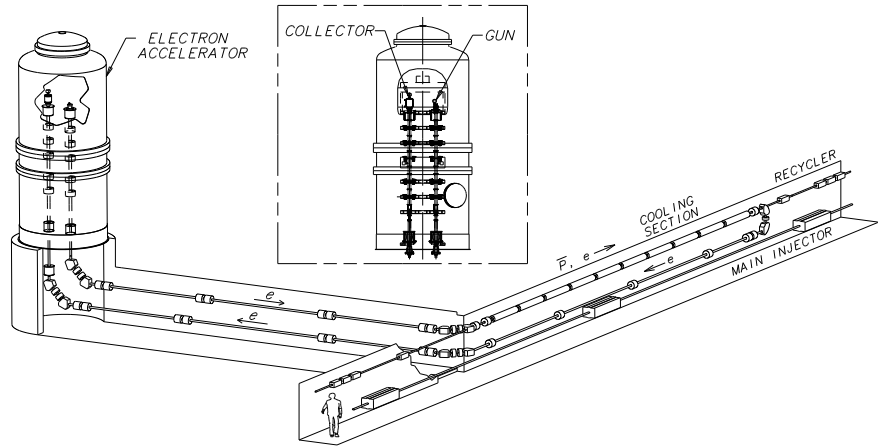

**Figure 6:** Schematic layout of the Recycler electron cooling system and accelerator cross-section (inset).

The Fermilab system employed a unique beam transport scheme [14]. The electron gun was immersed in a solenoidal magnetic field, which created a beam with large angular momentum. After the beam was extracted from the magnetic field and accelerated to 4.3 MeV, it was transported to the 20-m long cooling section solenoid using conventional focusing elements (as opposed to low-energy electron coolers where the beam remains immersed in a strong magnetic field at all times). The cooling section solenoid removed this angular momentum and the beam was made round and parallel such that the beam radius, $a$, produced the same magnetic flux, $Ba^2$, as at the cathode. The magnetic field in the cooling section was quite weak (100 G), therefore the kinetics of the electron-antiproton scattering was weakly affected by the magnetic field.

After becoming operational in September 2005, electron cooling in the Recycler allowed for significant improvements in Tevatron luminosity by providing higher intensity antiprotons with smaller emittances. With electron cooling, the Recycler has been able to store up to $6\times10^{12}$ antiprotons. In routine operations, the Recycler accumulated $3.5\text{-}4.0\times10^{12}$ antiprotons with a ~200-hr lifetime for injection to the Tevatron [15].

Among other unique features of the Recycler was the so-called barrier-bucket rf system [16] which allowed for crucial longitudinal beam manipulations of the accumulated antiproton beam.

## 5. Antiproton Flow and Tevatron Shot-Setup

As mentioned previously, stacks of freshly produced antiprotons were stored temporarily in the Accumulator. The Accumulator antiproton stack was periodically transferred to the Recycler where electron cooling allowed for a much larger antiproton intensity to be accumulated with smaller emittances. Typically 22-25$\times 10^{10}$ antiprotons were transferred to the Recycler every ~60 minutes. Prior to electron cooling in the Recycler, antiprotons destined for the Tevatron were extracted from the Accumulator only. Since late 2005, all Tevatron antiprotons were extracted from the Recycler only. Figure 7 illustrates the flow of antiprotons between the Accumulator, Recycler and Tevatron over a 1 week period.

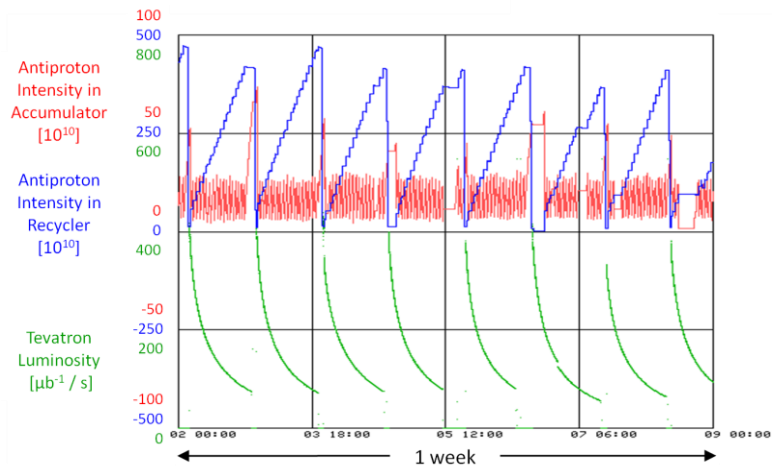

**Figure 7:** Production and transfers of antiprotons between the Accumulator and Recycler over 1 week of operation. While the Tevatron has a colliding beam store, small stacks of antiprotons are produced and stored in the Accumulator, and then periodically transferred to the Recycler in preparation for the subsequent Tevatron fill.

A typical Tevatron collider fill cycle is shown in Figure 8 [17]. First, proton bunches were injected two at a time on the central orbit. Then, electrostatic separators were powered to put the protons onto a helical orbit. Antiproton bunches were then injected (four bunches at a time) into gaps between the three proton bunch trains. After each group of 3 antiproton transfers, the gaps were cleared for the subsequent set of transfers by "cogging" the antiprotons – changing the antiproton RF cavity frequency to let them slip longitudinally relative to the protons. Once the beam loading was complete, the beams were accelerated to the top energy (in 86 seconds) and the machine optics was changed to the collision configuration in 25 steps over 125 sec (low-beta squeeze). The last two stages included initiating collisions at the two collision points and removing halo by moving in the collimators. The experiments then commenced data acquisition for the duration of the high-energy physics (HEP) store.

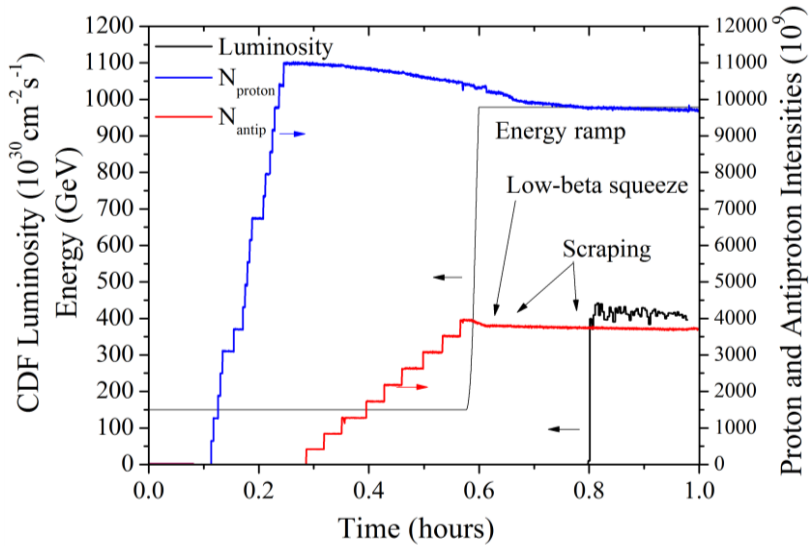

**Figure 8:** The collider fill cycle for store #8709 (May 2011).

**Summary**

For more than 25 years (1985-2011) the Fermilab antiproton complex was the centerpiece of the Tevatron collider program [18] and provided antiprotons for other particle physics experiments [19]. The continued Tevatron luminosity increase was mainly due to a larger number of antiprotons being available, which in turn was the result of a continuous and dedicated effort of hundreds of experts to optimize and improve antiproton accumulation and cooling. The antiproton stochastic and electron cooling methods were not invented at Fermilab, but they were perfected to a degree not achieved anywhere else in the world.